\documentclass[twocolumn,showpacs,amsmath,amssymb]{revtex4}
\usepackage{graphicx}
\usepackage{dcolumn}
\usepackage{bm}
\usepackage{slashed}
\usepackage{amsmath}
\usepackage{float}
\usepackage{subfig}

\def\be{\begin{equation}}
\def\ee{\end{equation}}
\def\ba{\begin{array}}
\def\ea{\end{array}}
\def\bea{\begin{eqnarray}}
\def\eea{\end{eqnarray}}

\begin{document}
\title{Analysis of strong decays of charmed
mesons $D^{*}_{2}(2460)$, $D_{0}(2560)$,  $D_{2}(2740)$,
$D_{1}(3000)$, $D^{*}_{2}(3000)$ and their spin partners
$D^{*}_{1}(2680)$, $D^{*}_{3}(2760)$ and $D^{*}_{0}(3000)$.}

\author{\large  Pallavi Gupta}
\author{\large A. Upadhyay}
\affiliation{School of Physics and Materials Science, Thapar
University, Patiala - 147004, Punjab, INDIA}
\date{\today}
\baselineskip =1\baselineskip

\begin{abstract}
Using the effective Lagrangian approach, we examine the recently
observed charm states $D^{*}_{J}(2460)$, $D_{J}(2560)$,
$D_{J}(2740)$, $D_{J}(3000)$ and their spin partners
$D^{*}_{J}(2680)$, $D^{*}_{J}(2760)$ and $D^{*}_{J}(3000)$ with
$J^{P}$ states $1P_{\frac{3}{2}}2^{+}$, $2S_{\frac{1}{2}}0^{-}$,
$1D_{\frac{5}{2}}2^{-}$, $2P_{\frac{1}{2}}1^{+}$ and
$2S_{\frac{1}{2}}1^{-}$, $1D_{\frac{5}{2}}3^{-}$,
$2P_{\frac{1}{2}}0^{+}$ respectively. We study their two body strong
decays, coupling constants and branching ratios with the emission of
light pseudo-scalar mesons $(\pi, \eta, K)$. We also analyze the
newly observed charm state $D^{*}_{2}(3000)$ and suggest it to be
either $1F(2^{+})$ or $2P(2^{+})$ state and justify one of them to
be the most favorable assignment for $D^{*}_{2}(3000)$.  We study
the partial and the total decay width of unobserved states $D(1
^{1}F_{3})$, $D_{s}(1 ^{1}F_{3})$ and $D_{s}(1 ^{1}F_{2})$ as the
spin and the strange partners of the $D^{*}_{2}(3000)$ charmed
meson. The branching ratios and the coupling constants $g_{TH}$,
$\widetilde{g}_{HH}$, $g_{YH}$, $\widetilde{g}_{SH}$ and $g_{ZH}$
calculated in this work can be confronted with the future
experimental data.
\end{abstract}
 \maketitle

\section{Introduction}
The excitation spectrum of $(c\overline{q})$ heavy-light charmed
mesons have received considerable theoretical and experimental
attention, as it provide opportunities to study the QCD properties
within the context of different models. Recently, LHCb collaboration
have used the Dalitz plot analysis to study the resonant
substructures $B^{-} \rightarrow D^{+}\pi^{-}\pi^{-}$ decays in the
pp collision at a center-of-mass energy 7 TeV. The masses and the
widths of charm resonances with spins 1, 2 and 3 at high
$D^{+}\pi^{-}$ masses are determined \cite{1}. The study gives
indication that, these resonances are mainly coming from the
contribution of the $D^{*}_{2}(2460)$,
 $D^{*}_{1}(2680)$, $D^{*}_{3}(2760)$ and $D^{*}_{2}(3000)$ charmed mesons. The measured Breit-Wigner masses and
widths of these charmed mesons are

\begin{multline}
 D^{*}_{2}(2460): M = 2463.7\pm0.4\pm0.4\pm0.6
\text{MeV},\\
\Gamma = 47.0\pm 0.8\pm0.9\pm0.3 \text{MeV},
\end{multline}
\begin{multline}
D^{*}_{1}(2680): M = 2681.1\pm5.6\pm4.9\pm13.1
\text{MeV},\\\Gamma=186.7\pm8.5\pm8.6\pm8.2 \text{MeV},
 \end{multline}
 \begin{multline}
D^{*}_{3}(2760): M = 2775.5\pm4.5\pm4.5\pm4.7 \text{MeV},\\
 \Gamma =
95.3\pm9.6\pm7.9\pm33.1 \text{MeV},
\end{multline}
\begin{multline}
 D^{*}_{2}(3000): M = 3214\pm29\mp33\mp36 \text{MeV},\\\Gamma = 186\pm38\pm34\pm63 \text{MeV}
\end{multline}

In 2010 and 2013, a great achievement have been made by BaBar and
LHCb collaboration. LHCb collaboration observed two natural parity
resonances $D^{*}_{J}(2650)^{0}$, $D^{*}_{J}(2760)^{0}$ and two
unnatural parity resonances $D_{J}(2580)^{0}$ and $D_{J}(2740)^{0}$
by studying the $D^{+}\pi^{-}$, $D^{0}\pi^{+}$ and $D^{*+}\pi^{-}$
invariant mass spectra \cite{2}. Along with these states, LHCb has
also observed $D_{J}(3000)^{0}$ in the $D^{*+}\pi^{-}$ final state
and $D^{*}_{J}(3000)^{+}$ and $D^{*}_{J}(3000)^{0}$ in the
$D^{0}\pi^{+}$ and $D^{+}\pi^{-}$ mass spectra respectively. BaBar
collaboration in 2010, observed $D_{J}(2560)^{0}$,
$D_{J}(2600)^{0}$, $D_{J}(2600)^{+}$, $D_{J}(2750)^{0}$,
$D_{J}^{*}(2760)^{+}$ and $D_{J}^{*}(2760)^{0}$ in the inclusive
$e^{+}e^{-}\rightarrow c\overline{c}$ interaction \cite{3}. Masses
and the widths of charm states predicted by BaBar and LHCb are so
close, that they are considered to be in the same $J^{P}$ state.
Masses and widths of these charm states observed by various
collaborations are presented in Table \ref{tab:expt}.
\setlength{\tabcolsep}{0.09em} %
{\renewcommand{\arraystretch}{0.2}%
\begin{table*}{\normalsize
\renewcommand{\arraystretch}{1.0}
\tabcolsep 0.2cm \caption{\label{tab:expt}The experimental results
from LHCb(2016)\cite{1}, LHCb(2013)\cite{2} and BaBar(2010)\cite{3}
of non-strange charm mesons. Values corresponding to M: and
$\Gamma:$ represents mass and decay width of the states. All the
values are in MeV unit.}
\begin{tabular}{c|c|c|c|c}
 \hline
  Charm State& LHCb(2013)\cite{2} & BaBar(2010)\cite{3}& LHCb(2016)\cite{1}& Decay Channel\\
  \hline
$D^{*}_{2}(2460)$&&&M:$2463.7\pm0.4\pm0.4$&$D^{*+}\pi^{-}$\\
&&&$\Gamma:47.0\pm 0.8\pm0.9$&\\
\hline
$D^{*}_{J}(2650)^{0}$&M:$2649.2\pm3.5\pm3.5$&M:$2608.7\pm2.4\pm2.5$&M:$2681.1\pm5.6\pm4.9$&$D^{*+}\pi^{-}$\\
&$\Gamma:140.2\pm17.1\pm18.6$&$\Gamma:93\pm6\pm13$&$\Gamma:186.7\pm8.5\pm8.6$&\\
\hline
$D^{*}_{J}(2760)^{0}$&M:$2761.1\pm5.1\pm6.5$&M:$2763.3\pm2.3\pm2.3$&M:$2775.5\pm4.5\pm4.5$&$D^{*+}\pi^{-}$\\
&$\Gamma:74.4\pm3.4\pm37.0$&$\Gamma:60.9\pm5.1\pm3.6$&$\Gamma:5.3\pm9.6\pm7.9$&\\
 \hline
$D_{J}(2560)^{0}$&M:$2579.5\pm3.4\pm5.5$&M:$2539.4\pm4.5\pm6.8$&&$D^{*+}\pi^{-}$\\
&$\Gamma:177.4\pm17.8\pm46.0$&$\Gamma:130\pm12\pm13$&&\\
\hline
$D_{J}(2740)^{0}$&M:$2737.0\pm3.5\pm11.24$&M:$2752.4\pm1.7\pm2.7$&&$D^{*+}\pi^{-}$\\
&$\Gamma:73.2\pm13.4\pm25.0$&$\Gamma:71\pm6\pm11$&&\\
\hline
$D_{J}(3000)^{0}$&M:$2971.8\pm8.7$&&&$D^{*+}\pi^{-}$\\
&$\Gamma:188.1\pm44.8$&&&\\
\hline
$D^{*}_{J}(2760)^{0}$&M:$2760.1\pm1.1\pm3.7$&&&$D^{+}\pi^{-}$\\
&$\Gamma:74.4\pm3.4\pm19.1$&&&\\
\hline
$D^{*}_{J}(3000)^{0}$&M:$3008.1\pm4.0$&&&$D^{+}\pi^{-}$\\
&$\Gamma:110.5\pm11.5$&&&\\
\hline
$D^{*}_{2}(3000)$&&&M:$3214\pm29\pm33\pm36$&$D^{+}\pi^{-}$\\
&&&$\Gamma:186\pm38\pm34\pm63$&\\
 \hline
$D_{J}^{*}(2760)^{+}$&M:$2771.7\pm1.7\pm3.8$&&&$D^{0}\pi^{+}$\\
&$\Gamma:66.7\pm6.6\pm10.5$&&&\\
\hline
$D^{*}_{J}(3000)^{+}$&M:$3008.1$&&&$D^{0}\pi^{+}$\\
&$\Gamma:110.5$&&&\\
 \hline
\end{tabular}
}
\end{table*}

 It is very crucial to assign a proper $J^{P}$ to the heavy-light
system in a given spectra, as large amount of experimental
information like decay width, branching ratios and hyperfine
splitting are based on their $J^{P}$. Various theoretical models
have suggested different $J^{P}$ states to the observed charm
mesons. In this paper, we analyze the available theoretical and
experimental data on the excited charm states and specify their
proper $J^{P}$. In our analysis, we mentioned $D^{*}_{2}(2460)$ to
be the well established state having $J^{P} = 2^{+}$ in the charm
spectra \cite{4}. The information provided by Babar (2010) and LHCb
(2013) for the states $D^{*}_{J}(2680)$ and $D^{*}_{J}(2760)$ were
confirmed in 2016 by LHCb, which had provided their J values as 1
and 3 respectively. Theoretical study of these two states concluded
their $J^{P}$ to be $1^{-}$ for $n = 2$ S-wave and $3^{-}$ for $n =
1$ D wave respectively \cite{5,6,7,8,9}. States $D_{J}(2560)^{0}$
and $D_{J}(2740)^{0}$ being the spin partners of
$D^{*}_{J}(2680)^{0}$ and $D^{*}_{J}(2760)^{0}$, are assigned $J^{P}
= 0^{-}$ for S-wave ($n=2$) and  $2^{-}$ for D-wave ($n=1$)
respectively.
Higher charm states $D^{*}_{J}(3000)$ and $D_{J}(3000)$ were studied
by various models like $^{3}P_{0}$ model, heavy quark effective
theory, but their $J^{P'}s$ are not yet confirmed. Authors in
\cite{11} assigned $D^{*}_{J}(3000)$ as the $1F_{\frac{5}{2}}2^{+}$
or $1F_{\frac{7}{2}}4^{+}$ state and $D_{J}(3000)$ as the
$1F_{\frac{7}{2}}3^{+}$ or $2P_{\frac{1}{2}}1^{+}$ state, but
Ref.\cite{10} have suggested various other possibilities for the
$J^{P'}s$ of $(D^{*}_{J}(3000)), (D_{J}(3000))$ and concluded
$2P(0^{+},1^{+})$ to be the most favorable $nLJ^{P'}s$ in the charm
spectra by studying their branching ratio.

Now, the main interest of theorists is on the newly predicted
$D_{2}^{*}(3000)$ state, whose mass and decay width is comparable
with the former $D^{*}_{J}(3000)$ state. It is suggested by Zhi-Gang
Wang in Ref.\cite{12}, that the energy gap between
$D_{2}^{*}(3000)^{0}$ and $D_{J}^{*}(3000)^{0}$ is 206 MeV
($M_{D_{2}^{*}(3000)^{0}} - M_{D_{J}^{*}(3000)^{0}} = 206 MeV$),
which indicates them to be different particles. On the basis of the
charm masses predicted by relativistic quark model \cite{13}, Wang
suggested $D_{2}^{*}(3000)$ to be $1F_{\frac{5}{2}} 2^{+}$ state
\cite{5,13}. Using the $^{3}P_{0}$ model, they also suggested the
most plausible assignment of $D_{2}^{*}(3000)$ to be the $3
P_{\frac{3}{2}}2^{+}$ state, but then the other possibility like $2
F_{\frac{5}{2}}2^{+}$  may not be completely excluded \cite{14}.
Thus, the clear picture of the $J^{P}$ of $D_{2}^{*}(3000)$ is not
yet available. This unclear picture is the motivation for our
present work.

 On the basis of masses predicted by various theoretical
models \cite{8,13,15,16,17,18}, we assume the two most favorable
$J^{P}$ states for $D_{2}^{*}(3000)$ to be either $1F(2^{+})$ or
$2P(2^{+})$. $D_{2}^{*}(3000)$ is observed in the decay channel
$D^{+}\pi^{-}$ but not in $D^{*+}\pi^{-}$, and hence $D^{*+}\pi^{-}$
decay mode must be suppressed. By analyzing the branching ratio
BR=$\frac{\Gamma(D_{2}^{*}(3000)\rightarrow
D^{*}\pi)}{\Gamma(D_{2}^{*}(3000)\rightarrow D\pi)}$ with their
masses and strong decay widths, we further choose one of them as the
best possible $J^{P}$ state for the $D_{2}^{*}(3000)$ and have
determine its strong coupling constant. We use HQET model for
studying the decay widths at the leading order approximations,
because the mass and the spin degeneracy of heavy hadrons appears as
approximate internal symmetry of the Lagrangian. Beside the fact
that, HQET contains many unknown phenomenological constants, HQET in
conjugation with the chiral perturbation theory, has been
successfully applied to the strong decays of the heavy hadrons
\cite{18a,18b}. Heavy quark symmetry helps in reducing the
parameters by  imposing constraints on these constants, like the
range of the strong coupling constants is constrained to be with in
0 and 1 by studying the decay widths and branching ratios of ground
state charm mesons.\cite{18c}. The strong couplings can also be
retrieved by comparing the strong decay widths with the experimental
available decay widths and masses. The paper is arranged as follows:
section 2 gives the brief review of the HQET model(For the detailed
review refer Ref.s \cite{19,20,21,22} ). In section 3, we study the
strong decays and the branching ratios of the $D^{*}_{J}(2460)$,
$D_{J}(2560)$, $D_{J}(2740)$, $D_{J}(3000)$ and their spin partners
$D^{*}_{J}(2680)$, $D^{*}_{J}(2760)$ and $D^{*}_{J}(3000)$ with
$J^{P}$ states $1P_{\frac{3}{2}}2^{+}$, $2S_{\frac{1}{2}}0^{-}$,
$1D_{\frac{5}{2}}2^{-}$, $2P_{\frac{1}{2}}1^{+}$ and
$2S_{\frac{1}{2}}1^{-}$, $1D_{\frac{5}{2}}3^{-}$,
$2P_{\frac{1}{2}}0^{+}$ respectively and discusses their strong
coupling constants involved. We also analyze the newly observed
charm state $D^{*}_{2}(3000)$ and suggest it to be either
$1F(2^{+})$ or $2P(2^{+})$ state. And by studying the decay behavior
and the branching ratio for both these nL$J^{P}$'s, we justify one
of them to be the most favorable assignment for $D^{*}_{2}(3000)$.
In addition to this, we also study the strong decays for the
unobserved spin and the strange partners of $D^{*}_{2}(3000)$ i.e.
$D(1 ^{1}F_{3})$, $D_{s}(1 ^{1}F_{3})$ and $D_{s}(1 ^{1}F_{2})$ in
the framework of the HQET, which are experimentally unobserved but
theoretically predicted. Section 4 presents the conclusion of our
work.

\section{Framework}
 In the heavy quark limit $m_{Q}
>> \Lambda_{QCD} >> m_{q}$, Compton wave-length of the heavy quark $\lambda_{Q} \simeq 1/m_{Q}$ is much
 smaller than the hadronic distance 1fm. The strong interactions of such a heavy quark with light
 quarks and gluons can be described by an effective theory, which is
invariant with flavor and the spin of the heavy quark. This
effective theory involves the corrections at the order of $1/m_{Q}$
order. The theoretical framework for such analysis is provided by
the so-called heavy quark effective theory. Also, the mass and spin
degeneracy of the heavy hadrons appears as approximate internal
symmetries of the Lagrangian. It is an effective QCD theory for
$N_{f}$ heavy quarks Q with their four velocity fixed. In this
theory, spin and parity of the heavy quark decouples from the light
degrees of freedom as they interact through the exchange of soft
gluons. Heavy mesons are classified in doublets, in relation to the
total conserved angular momentum $i.e.$ $s_{l}=s_{\overline{q}}+l$,
where $s_{\overline{q}}$ and $\textit{l}$ are the spin and orbital
angular momentum
 of the light degree of freedom respectively. For $\textit{l} =0$ (S-wave), the
doublet is represented by $(P, P^{*})$ with $J^{P}_{s_{l}}=
(0^{-},1^{-})_{\frac{1}{2}}$, which for $\textit{l} =1$ (P-wave),
there are two doublets represented by $(P^{*}_{0},P^{'}_{1})$ and
$(P_{1},P^{*}_{2})$ with $J^{P}_{s_{l}}=(0^{+},1^{+})_{\frac{1}{2}}$
and $(1^{+},2^{+})_{\frac{3}{2}}$ respectively. Two doublets of
$\textit{l}=2$ (D-wave) are represented by $(P^{*}_{1},P_{2})$ and
$(P_{2}^{'},P^{*}_{3})$ belonging to
$J^{P}_{s_{l}}=(1^{-},2^{-})_{\frac{3}{2}}$ and
$(2^{-},3^{-})_{\frac{5}{2}}$ respectively. And the doublets of
$\textit{l}=3$ (F-wave) are represented by $(P^{*}_{2},P_{3})$ and
$(P_{3}^{'},P^{*}_{4})$ for
$J^{P}_{s_{l}}=(2^{+},3^{+})_{\frac{5}{2}}$ and
$(3^{+},4^{+})_{\frac{7}{2}}$ respectively. These doublets are
described by the effective super-field $H_{a}, S_{a}, T_{a}, X_{a},
Y_{a}$ and $Z_{a}$ \cite{27,23}.
\begin{gather}
\label{eq:lagrangian}
 H_{a}=\frac{1+\slashed
v}{2}\{P^{*}_{a\mu}\gamma^{\mu}-P_{a}\gamma_{5}\}\\
S_{a}=\frac{1+\slashed
v}{2}\{P^{\mu}_{1a}\gamma_{\mu}\gamma_{5}-P^{*}_{0a}\}\\
T^{\mu}_{a}=\frac{1+\slashed v}{2}
\{P^{*\mu\nu}_{2a}\gamma_{\nu}-P_{1a\nu}\sqrt{\frac{3}{2}}\gamma_{5}
[g^{\mu\nu}-\frac{\gamma^{\nu}(\gamma^{\mu}-\upsilon^{\mu})}{3}]\}
\end{gather}
\begin{multline}
Y^{\mu\nu}_{a}=\frac{1+\slashed
v}{2}\{P^{*\mu\nu\sigma}_{3a}\gamma_{\sigma}-P^{\alpha\beta}_{2a}\sqrt{\frac{5}{3}}\gamma_{5}\\
[g^{\mu}_{\alpha}g^{\nu}_{\beta}-
\frac{g^{\nu}_{\beta}\gamma_{\alpha}(\gamma^{\mu}-v^{\mu})}{5}-\frac{g^{\mu}_{\alpha}\gamma_{\beta}(\gamma^{\nu}-v^{\nu})}{5}]\}
\end{multline}
\begin{multline}
Z^{\mu\nu}_{a}=\frac{1+\slashed
v}{2}\{P^{\mu\nu\sigma}_{3a}\gamma_{5}\gamma_{\sigma}-P^{*\alpha\beta}_{2a}\sqrt{\frac{5}{3}}\\
[g^{\mu}_{\alpha}g^{\nu}_{\beta}-
\frac{g^{\nu}_{\beta}\gamma_{\alpha}(\gamma^{\mu}+v^{\mu})}{5}-\frac{g^{\mu}_{\alpha}\gamma_{\beta}(\gamma^{\nu}+v^{\nu})}{5}]\}
\end{multline}

Here the field $H_{a}$ describe the $(P,P^{*})$ doublet i.e. S-wave,
$S_{a}$ and $T_{a}$ fields represents the P-wave doublets
$(0^{+},1^{+})_{\frac{1}{2}}$ and $(1^{+},2^{+})_{\frac{3}{2}}$
respectively. The mentioned indices a or b in the subsequent fields
and Lagrangian are $SU(3)$ flavor index (u, d or s). P and $P^{*}$
in field $ H_{a}$ represents $D^{0}, D^{+}, D^{+}_{s}$ and $D^{*0},
D^{*+}, D^{*+}_{s}$ respectively. The heavy meson field $P^{(*)}$
contain a factor $\sqrt{m_{Q}}$ with mass dimension of
$\frac{1}{2}$. For the radially excited states with radial quantum
number n=2, these states are replaced by $\widetilde{P},
\widetilde{P}^{*}$ and so on. The properties of the hadrons are
invariant under $SU(2N_{f})$ transformations, hence heavy quark spin
and flavor symmetries provide a clear picture for the study of the
heavy-light mesons in heavy quark physics. The light pseudoscalar
mesons are described by the fields $\xi=
exp^{\frac{i\mathcal{M}}{f_{\pi}}}$, where $\mathcal{M}$ is defined
as
\begin{center}
\begin{equation}
\mathcal{M} = \begin{pmatrix}
\frac{1}{\sqrt{2}}\pi^{0}+\frac{1}{\sqrt{6}}\eta & \pi^{+} & K^{+}\\
\pi^{-} & -\frac{1}{\sqrt{2}}\pi^{0}+\frac{1}{\sqrt{6}}\eta &
K^{0}\\
K^{-} & \overline{K}^{0} & -\sqrt{\frac{2}{3}}\eta
\end{pmatrix}
\end{equation}
\end{center}
The pion octet is introduced by the vector and axial vector
combinations
$V^{\mu}=\frac{1}{2}(\xi\partial^{\mu}\xi^{\dag}+\xi^{\dag}\partial^{\mu}\xi)$
and
$A^{\mu}=\frac{1}{2}(\xi\partial^{\mu}\xi^{\dag}-\xi^{\dag}\partial^{\mu}\xi)$.
We choose $f_{\pi}=130MeV$. Here, all traces are taken over Dirac
spinor indices, light quark $SU(3)_{V}$ flavor indices a = u, d, s
and heavy quark flavor indices Q = c, b. The Dirac structure of the
chiral Lagrangian is given by the velocity vector v/c. At the
leading order approximation, the heavy meson chiral lagrangians
$L_{HH}$, $L_{SH}$, $L_{TH}$, $L_{YH}$, $L_{ZH}$ for
 the two-body strong interactions through light pseudoscalar mesons
 are written as :
\begin{center}
\begin{gather}
\label{eq:lagrangian}
 L_{HH}=g_{HH}Tr\{\overline{H}_{a}
 H_{b}\gamma_{\mu}\gamma_{5}A^{\mu}_{ba}\}\\
L_{SH}=g_{SH}Tr\{\overline{H}_{a}S_{b}\gamma_{\mu}\gamma_{5}A^{\mu}_{ba}\}+h.c.\\
L_{TH}=\frac{g_{TH}}{\Lambda}Tr\{\overline{H}_{a}T^{\mu}_{b}(iD_{\mu}\slashed
A + i\slashed D A_{\mu})_{ba}\gamma_{5}\}+h.c.
\end{gather}
\begin{multline}
L_{YH}=\frac{1}{\Lambda^{2}}Tr\{\overline{H}_{a}Y^{\mu\nu}_{b}[k^{Y}_{1}\{D_{\mu}
,D_{\nu}\}A_{\lambda}+k^{Y}_{2}(D_{\mu}D_{\lambda}A_{\nu}\\
+D_{\nu}D_{\lambda}A_{\mu})]_{ba}\gamma^{\lambda}\gamma_{5}\}+h.c.
\end{multline}
\begin{multline}
L_{ZH}=\frac{1}{\Lambda^{2}}Tr\{\overline{H}_{a}Z^{\mu\nu}_{b}[k^{Z}_{1}\{D_{\mu}
,D_{\nu}\}A_{\lambda}+\\
k^{Z}_{2}(D_{\mu}D_{\lambda}A_{\nu}+D_{\nu}D_{\lambda}A_{\mu})]_{ba}\gamma^{\lambda}\gamma_{5}\}+h.c.
\end{multline}
\end{center}
In these equations $D_{\mu} =
\partial_{\mu}+V_{\mu}$,  $\{D_{\mu},D_{\nu}\}
= D_{\mu}D_{\nu}+D_{\nu}D_{\mu}$ and $\{D_{\mu} ,D_{\nu}D_{\rho}\} =
D_{\mu}D_{\nu}D_{\rho}+D_{\mu}D_{\rho}D_{\nu}+D_{\nu}D_{\mu}D_{\rho}+D_{\nu}D_{\rho}D_{\mu}+D_{\rho}D_{\mu}
D_{\nu}+D_{\rho}D_{\nu}D_{\mu}$. $\Lambda$ is the chiral symmetry
breaking scale taken as 1 GeV. $g_{HH}$, $g_{SH}$, $g_{TH}$, $g_{YH}
= k^{Y}_{1}+k^{Y}_{2}$ and $g_{ZH} = k^{Z}_{1}+k^{Z}_{2}$ are the
strong coupling constants involved. The above equations describe the
interactions of higher excited charm states to the ground state
positive and negative parity charm mesons along with the emission of
light pseudo-scalar mesons $(\pi,\eta,K)$. Using the lagrangians
$L_{HH}, L_{SH}, L_{TH}, L_{YH}, L_{ZH}$, the two body strong decays
of $Q\overline{q}$ heavy-light charm mesons are given as
\\$(0^{-},1^{-}) \rightarrow (0^{-},1^{-}) + M$
\begin{gather}
\label{eq:lagrangian} \Gamma(1^{-} \rightarrow 1^{-})=
C_{M}\frac{g_{HH}^{2}M_{f}p_{M}^{3}}{3\pi f_{\pi}^{2}M_{i}}\\
\Gamma(1^{-} \rightarrow 0^{-})=
C_{M}\frac{g_{HH}^{2}M_{f}p_{M}^{3}}{6\pi f_{\pi}^{2}M_{i}}\\
\Gamma(0^{-} \rightarrow 1^{-})=
C_{M}\frac{g_{HH}^{2}M_{f}p_{M}^{3}}{2\pi f_{\pi}^{2}M_{i}}
\end{gather}

 $(0^{+},1^{+}) \rightarrow (0^{-},1^{-}) + M$
\begin{gather}
\label{eq:lagrangian} \Gamma(1^{+} \rightarrow 1^{-})=
C_{M}\frac{g_{SH}^{2}M_{f}(p^{2}_{M}+m^{2}_{M})p_{M}}{2\pi f_{\pi}^{2}M_{i}}\\
\Gamma(0^{+} \rightarrow 0^{-})=
C_{M}\frac{g_{SH}^{2}M_{f}(p^{2}_{M}+m^{2}_{M})p_{M}}{2\pi
f_{\pi}^{2}M_{i}}
\end{gather}

 $(1^{+},2^{+}) \rightarrow (0^{-},1^{-}) + M$
\begin{gather}
\label{eq:lagrangian} \Gamma(2^{+} \rightarrow 1^{-})=
C_{M}\frac{2g_{TH}^{2}M_{f}p_{M}^{5}}{5\pi f_{\pi}^{2}\Lambda^{2}M_{i}}\\
\Gamma(2^{+} \rightarrow 0^{-})=
C_{M}\frac{4g_{TH}^{2}M_{f}p_{M}^{5}}{15\pi f_{\pi}^{2}\Lambda^{2}M_{i}}\\
\Gamma(1^{+} \rightarrow 1^{-})=
C_{M}\frac{2g_{TH}^{2}M_{f}p_{M}^{5}}{3\pi
f_{\pi}^{2}\Lambda^{2}M_{i}}
\end{gather}


$(2^{-},3^{-}) \rightarrow (0^{-},1^{-}) + M$

\begin{gather}
\label{eq:lagrangian} \Gamma(2^{-} \rightarrow 1^{-})=
C_{M}\frac{4g_{YH}^{2}}{15\pi f_{\pi}^{2}\Lambda^{4}}
\frac{M_{f}}{M_{i}}[p_{M}^{7}]\\
\Gamma(3^{-} \rightarrow 0^{-})= C_{M}\frac{4g_{YH}^{2}}{35\pi
f_{\pi}^{2}\Lambda^{4}} \frac{M_{f}}{M_{i}}[p_{M}^{7}]\\
\Gamma(3^{-} \rightarrow 1^{-})= C_{M}\frac{16g_{YH}^{2}}{105\pi
f_{\pi}^{2}\Lambda^{4}} \frac{M_{f}}{M_{i}}[p_{M}^{7}]
\end{gather}

$(2^{+},3^{+}) \rightarrow (0^{-},1^{-}) + M$

\begin{gather}
\label{eq:lagrangian} \Gamma(2^{+} \rightarrow 1^{-})=
C_{M}\frac{8g_{ZH}^{2}}{75\pi f_{\pi}^{2}\Lambda^{4}}
\frac{M_{f}}{M_{i}}[p_{M}^{5}(m_{M}^{2}+p_{M}^{2})]\\
 \Gamma(2^{+} \rightarrow 0^{-})=
C_{M}\frac{4g_{ZH}^{2}}{25\pi f_{\pi}^{2}\Lambda^{4}}
\frac{M_{f}}{M_{i}}[p_{M}^{5}(m_{M}^{2}+p_{M}^{2})]\\
\Gamma(3^{+} \rightarrow 1^{-})= C_{M}\frac{4g_{ZH}^{2}}{25\pi
f_{\pi}^{2}\Lambda^{4}}
\frac{M_{f}}{M_{i}}[p_{M}^{5}(m_{M}^{2}+p_{M}^{2})]
\end{gather}


In the above decay widths, $M_{i}$ and $M_{f}$ stands for initial
and final meson mass, $p_{M}$ and $m_{M}$ are the final momentum and
mass of the light pseudo-scalar meson respectively. The coefficient
$C_{\pi^{\pm}}, C_{K^{\pm}}, C_{K^{0}}, C_{\overline{K}^{0}}=1$,
$C_{\pi^{0}}=\frac{1}{2}$ and $C_{\eta}=\frac{2}{3}$ or
$\frac{1}{6}$. Different values of $C_{\eta}$ corresponds to the
initial state being $c\overline{u}, c\overline{d}$ or $
c\overline{s}$ respectively. All hadronic coupling constants depends
on the radial quantum number. For the decay within n=1 they are
notated as $g_{HH}$, $g_{SH}$ etc, and the decay from n=2 to n=1
they are represented by $\widetilde{g}^{2}_{HH}$,
$\widetilde{g}^{2}_{SH}$,
 Higher order corrections for spin and
flavor violation of order $\frac{1}{m_{Q}}$ are excluded to avoid
new unknown coupling constants. Equations 16-29 shows that the decay
width of any state depends on the initial and final meson masses,
their strong coupling constants, pion decay constant, energy scale
$\Lambda$, mass and momentum of light pseudo-scalar mesons. Unknown
coupling constants in these widths, can either be theoretically
predicted or can be determined indirectly from the known
experimental values of the decay widths. Theoretically, lattice QCD
\cite{25}, QCD sum rules \cite{26} have successfully predicted some
of these coupling constants. The numerical masses of various mesons
used in the calculation are listed in Table \ref{tab:input}.

\setlength{\tabcolsep}{0.09em} %
{\renewcommand{\arraystretch}{0.2}%

\begin{table*}{\normalsize
\renewcommand{\arraystretch}{1.0}
\tabcolsep 0.2cm \caption{\label{tab:input}Numerical value of the
meson masses used in this work \cite{4}.}
 \noindent
\begin{tabular}{ccccccc}
  \hline
  \hline
  States &$D^{0}$&$D^{\pm}$&$D^{*+}$&$D^{*0}$&$D_{S}^{+}$&$D^{*+}_{S}$\\
  \hline
  Masses(MeV)&1864.86&1869.62&2010.28&2006.98&1968.49&2112.30\\
  \hline
  States &$\pi^{\pm}$&$\pi^{0}$&$\eta$&$K^{+}$&$K^{0}$&\\
  \hline
  Masses(MeV)&139.57&134.97&547.85&493.67&497.61&\\
\hline \hline
\end{tabular}}
\end{table*}
\section{Numerical Analysis}
Assigning a proper $J^{P'}s$ to the experimentally available states
are essential, as it helps in retrieving many properties like decay
width, strong coupling constant, branching ratios etc of these
states. In this paper, we reanalyze the previously available
theoretical and experimental data on the charm states
$D^{*}_{J}(2460)$, $D_{J}(2560)$, $D_{J}(2740)$,
 $D^{*}_{J}(2680)$, $D^{*}_{J}(2760)$, $D_{J}(3000)$ and
 $D^{*}_{J}(3000)$. This analysis is based on the available
 information on J values taken from LHCb in 2016. Hence we identify these states as:
\begin{center}
\begin{gather}
\label{eq:lagrangian}
 D^{*}_{J}(2460) = (2^{+})_{\frac{3}{2}} \text{with  }  n=1, L=1,\\
(D_{J}(2560), D^{*}_{J}(2680)) = (0^{-}, 1^{-})_{\frac{1}{2}}
 \text{with  }n=2, L=0,\\
(D_{J}(2740), D^{*}_{J}(2760)) = (2^{-}, 3^{-})_{\frac{5}{2}}
 \text{with  } n=1, L=2,\\
D^{*}_{J}(3000)), (D_{J}(3000) = (0^{+}, 1^{+})_{\frac{1}{2}}
 \text{with  } n=2, L=1
\end{gather}
\end{center}
The numerical value of the partial decay widths and the ratios for
the charm states $D^{*}_{2}(2460)$, $D_{0}(2560)$, $D_{2}(2740)$,
 $D^{*}_{1}(2680)$, $D^{*}_{3}(2760)$, $D_{1}(3000)$ and $D^{*}_{0}(3000)$ are
listed in Table \ref{tab:width}. We equate the calculated decay
widths  with the experimental data in Table \ref{tab:width} to
obtain the coupling constants which are listed in Table
\ref{tab:coup}. The couplings
$\widetilde{g}_{HH},\widetilde{g}_{SH}$ are obtained by averaging
the values obtained from ($D_{0}(2560)$, $D^{*}_{1}(2680)$) and
($D_{1}(3000)$, $D^{*}_{0}(3000)$) respectively. We have neglected
the small value of the coupling $g_{YH} = 0.10$, in comparison with
its other theoretically predicted values \cite{23}. The range in the
coupling constant, comes from the error-bar in the experimental mass
and decay width values.

\setlength{\tabcolsep}{0.09em} %
{\renewcommand{\arraystretch}{0.2}%

\begin{table*}{\normalsize
\renewcommand{\arraystretch}{1.0}
\tabcolsep 0.2cm \caption{Strong decay width of newly observed charm
mesons $D^{*}_{2}(2460)$, $D_{0}(2560)$, $D_{2}(2740)$,
 $D^{*}_{1}(2680)$, $D^{*}_{3}(2760)$, $D_{1}(3000)$ and
 $D^{*}_{0}(3000)$. Ratio in 5th column represents the $\widehat{{\bf \Gamma}}=
\frac{\Gamma}{\Gamma(D_{J}^{*} \rightarrow D^{*+}\pi^{-})}$ for the
mesons. Fraction gives the percentage of the partial decay width
with respect to the total decay width. }\label{tab:width}
\begin{center}
\begin{tabular}{c|c|c|c|c|c|c}
\hline \hline State&$nLs_{l}J^{P}$&Decay channel&Decay
Width(MeV)&Ratio&Fraction&Experimental value(MeV)\\
\hline
$D^{*}_{2}(2460)$&1$P_{3/2}2^{+}$&$D^{*+}\pi^{-}$&$56.55g^{2}_{TH}$&1&20.05&\\
&&$D^{*+}\pi^{0}$&$29.76g^{2}_{TH}$&0.52&10.55&\\
&&$D^{*+}\eta$&-&-&0&\\
&&$D^{+}\pi^{-}$&$128.40g^{2}_{TH}$&2.27&45.52&\\
&&$D^{+}\pi^{0}$&$67.06g^{2}_{TH}$&1.18&23.77&\\
&&$D^{+}\eta$&0.26$g^{2}_{TH}$&0&0&\\
&&Total&282.04$g^{2}_{TH}$&&&$47.00\pm0.80$
 \text{\cite{1}}\\
 \hline
$D_{0}(2560)$&$2S_{1/2}0^{-}$&$D^{*+}\pi^{-}$&$867.32\widetilde{g}^{2}_{HH}$&1&65.99&\\
&&$D^{*+}\pi^{0}$&$443.03\widetilde{g}^{2}_{HH}$&0.51&33.71&\\
&&$D^{*+}\eta$&$3.858\widetilde{g}^{2}_{HH}$&0&0.29&\\
&&Total&1314.22$\widetilde{g}^{2}_{HH}$&&&$177.40\pm17.80 \text{\cite{2}}$\\
\hline
$D^{*}_{1}(2680)$&2$S_{1/2}1^{-}$&$D^{*+}\pi^{-}$&$889.34\widetilde{g}^{2}_{HH}$&1&32.41&\\
&&$D^{*+}\pi^{0}$&4$451.87\widetilde{g}^{2}_{HH}$&0.50&16.56&\\
&&$D^{*+}\eta$&$31.07\widetilde{g}^{2}_{HH}$&0.03&1.13&\\
&&$D^{*+}_{s}K^{-}$&$78.40\widetilde{g}^{2}_{HH}$&0.08&2.87&\\
&&$D^{+}\pi^{-}$&$682.53\widetilde{g}^{2}_{HH}$&0.76&25.01&\\
&&$D^{+}\pi^{0}$&$346.56\widetilde{g}^{2}_{HH}$&0.38&12.70&\\
&&$D^{+}\eta$&$48.05\widetilde{g}^{2}_{HH}$&0.05&1.76&\\
&&$D^{+}_{s}K^{-}$&$200.49\widetilde{g}^{2}_{HH}$&0.22&7.34&\\
&&Total&2728.35$\widetilde{g}^{2}_{HH}$&&&$186.70\pm8.50  \text{\cite{1}}$\\
 \hline
$D_{2}(2740)$&$1D_{5/2}2^{-}$&$D^{*+}\pi^{-}$&$127.35g^{2}_{YH}$&1&64.79&\\
&&$D^{*+}\pi^{0}$&$65.96g^{2}_{YH}$&0.51&33.55&\\
&&$D^{*+}\eta$&$1.30g^{2}_{YH}$&0.01&0.97&\\
&&$D^{*+}_{s}K^{-}$&$1.92g^{2}_{YH}$&0.01&0.97&\\
&&Total&196.55$g^{2}_{YH}$&&&$73.20\pm13.40$  \text{\cite{2}}\\
\hline
$D^{*}_{3}(2760)$&1$D_{5/2}3^{-}$&$D^{*+}\pi^{-}$&$100.15g^{2}_{YH}$&1&21.10&\\
&&$D^{*+}\pi^{0}$&$51.73g^{2}_{YH}$&0.51&10.90&\\
&&$D^{*+}\eta$&$1.53g^{2}_{YH}$&0.01&0.32&\\
&&$D^{*+}_{s}K^{-}$&$2.88g^{2}_{YH}$&0.02&0.60&\\
&&$D^{+}\pi^{-}$&$191.14g^{2}_{YH}$&1.90&40.28&\\
&&$D^{+}\pi^{0}$&$98.82g^{2}_{YH}$&0.98&20.82&\\
&&$D^{+}\eta$&$7.05g^{2}_{YH}$&0.07&1.48&\\
&&$D^{+}_{s}K^{-}$&$21.14g^{2}_{YH}$&0.21&4.45&\\
&&Total&474.47$g^{2}_{YH}$&&&$95.30\pm9.60 \text{\cite{1}}$\\
\hline
$D_{1}(3000)$&$2P_{1/2}1^{+}$&$D^{*+}\pi^{-}$&$3325.52\widetilde{g}^{2}_{SH}$&1&41.96&\\
&&$D^{*+}\pi^{0}$&$1674.26\widetilde{g}^{2}_{SH}$&0.50&21.12&\\
&&$D^{*+}\eta$&$516.82\widetilde{g}^{2}_{SH}$&0.15&6.52&\\
&&$D^{*+}_{s}K^{-}$&$2408.76\widetilde{g}^{2}_{SH}$&0.72&30.39&\\
&&Total&7925.36$\widetilde{g}^{2}_{SH}$&&&$188.10\pm44.60  \text{\cite{2}}$\\
\hline
 $D^{*}_{0}(3000)$&$2P_{1/2}0^{+}$&$D^{+}\pi^{-}$&$2315.81\widetilde{g}^{2}_{SH}$&0.50&20.26&\\
&&$D^{+}\pi^{0}$&$4598.65\widetilde{g}^{2}_{SH}$&1&40.24&\\
&&$D^{+}\eta$&$748.382\widetilde{g}^{2}_{SH}$&0.16&6.54&\\
&&$D^{+}_{s}K^{-}$&$3763.23\widetilde{g}^{2}_{SH}$&0.81&32.93&\\
&&Total&11426.10$\widetilde{g}^{2}_{SH}$&&&$110.50\pm11.50 \text{\cite{2}}$\\
\hline \hline
\end{tabular}
\end{center}}
\end{table*}
}

\setlength{\tabcolsep}{0.09em} %
{\renewcommand{\arraystretch}{0.2}%

\begin{table*}{\normalsize
\renewcommand{\arraystretch}{1.0}
\tabcolsep 0.2cm \caption{Value of various coupling constants
obtained in the literature.}\label{tab:coup}
\begin{center}
\begin{tabular}{c|c|c|c}
\hline
Coupling constant&Our calculation&Work in \cite{23}&Work in \cite{9}\\
\hline
$g_{TH}$&$0.40\pm0.01$&$0.43\pm0.05$&$0.43\pm0.01$\\
$\widetilde{g}_{HH}$&$0.31\pm0.05$&$0.14\pm0.03$&$0.28\pm0.01$\\
$g_{YH}$&$0.61\pm0.05$&$0.53\pm0.13$&$0.42\pm0.02$\\
$\widetilde{g}_{SH}$&$0.12\pm0.03$&-&-\\
\hline
\end{tabular}
\end{center}}
\end{table*}
 On the basis of the theoretically predicted masses
\cite{8,13,15,16,17,18}, $D_{2}^{*}(3000)$ is assumed to belong to
either $1F_{\frac{5}{2}}(2^{+})$ or $2P_{\frac{3}{2}}(2^{+})$ state.
The partial and the total decay widths for both these states are
shown in Table \ref{tab:two}. To clear out the $J^{P}$ state for
$D_{2}^{*}(3000)$ between $1F(2^{+})$ and $2P(2^{+})$, we have
observed the BR=$\frac{\Gamma(D_{2}^{*}(3000)\rightarrow
D^{*}\pi)}{\Gamma(D_{2}^{*}(3000)\rightarrow D\pi)}$ for both these
states with their masses. The graph for the BR with the masses for
the two $J^{P}$ states are shown in Figure 1. The graph 1(a) shows,
the value of BR for $2P_{\frac{3}{2}}(2^{+})$ is equal to 1.06
corresponding to the mass 3214 MeV, predicting $D^{*}\pi$ to be
dominant mode as compared to $D\pi$. And the graph 1(b) depicts the
value of BR for $1F_{\frac{5}{2}}(2^{+})$ state to be 0.40 for mass
3214 MeV, predicting $D\pi$ to be the dominant mode. Since the
$D^{*}\pi$ decay channel for $D_{2}^{*}(3000)$ is experimentally
suppressed, therefore $1F(2^{+})$ is considered to be the most
favorable $J^{P}$ for $D_{2}^{*}(3000)$.

Along with the decay channels mentioned in Table \ref{tab:two},
$D_{2}^{*}(3000)$ being $1F(2^{+})$ also decays to $1 P(1^{+})$, $1
P^{'}(1^{+})$, $1 D(2^{-})$ and $1 D^{'}(2^{-})$ states along with
pseudoscalar mesons $(\pi,\eta,K)$. Since these decays occur via
relative F-wave and D-wave, the contribution of their phase space to
the decay widths are negligible. And therefore, these channels are
suppressed. Considering the decay channels mentioned in
Table\ref{tab:two} to be the only dominant decay modes,
 the total decay width of $D_{2}^{*}(3000)$ comes out to be
7690.53$g^{2}_{ZH}$. Along with the partial decay widths, Table
\ref{tab:two} shows the ratio $\widehat{{\bf \Gamma}}=
\frac{\Gamma}{\Gamma(D_{2}^{*}(3000) \rightarrow D^{*+}\pi^{-})}$
and the branching fraction for the decay channels of
$D_{2}^{*}(3000)$ state. The results in Table \ref{tab:two} reveals
that,  for $D_{2}^{*}(3000)$ state $D^{+}\pi^{-}$ and $D^{0}\pi^{0}$
are the main decay modes as compared to the $D^{*+}\pi^{-}$ mode.
The decay width obtained in this work is finally compared with the
experimental result, and the coupling constant $g_{ZH}$ is obtained
as
\begin{equation}
 g_{ZH} = 0.15\pm0.02
\end{equation}
The information on the value of coupling $g_{ZH}$ is very limited in
the literature, so extracting its value will be useful for the
theory, in finding partial and the total decay widths of unobserved
charm states  $D(1 ^{1}F_{3})$, $D_{s}(1 ^{1}F_{3})$ and $D_{s}(1
^{3}F_{2})$. Until now, the experimental information on the strong
decay widths of $D(1 ^{1}F_{3})$, $D_{s}(1 ^{1}F_{3})$ and $D_{s}(1
^{3}F_{2})$ states is unavailable, so the prediction of their
partial and total decay widths will be a motivation for future
experiments. Mass of $D(1 ^{1}F_{3})$ is predicted to be $3099\pm25$
MeV Ref.\cite{13,16,17,18}. OZI allowed decay channels of $D(1
^{1}F_{3})$ are listed in the Table \ref{tab:width1f}. Column 4 of
the Table \ref{tab:width1f} gives the ratio of the partial decay
widths for $D(1 ^{1}F_{3})$ with respect to its partial decay width
$D^{*+}\pi^{-}$.
 Apart from the decay channels listed in Table \ref{tab:width1f}, $D(1 ^{1}F_{3})$ also decays to P-wave
charm meson states through the light pseudo-scalar meson, the decay
occurs via. F-wave, and due to small phase space, these modes are
suppressed and not considered in the present work. From the listed
decay channels, $D^{*+}\pi^{-}$ comes out to be the dominant decay
mode for $D(1 ^{1}F_{3})$ with branching fraction $51.84\%$. Hence,
the decay channel $D^{*+}\pi^{-}$ is suitable for the experimental
search for the missing charm state $D(1 ^{1}F_{3})$ in future. Using
the value of the coupling constant $g_{ZH}$ obtained from equation
34, the total decay width of the charm state $D(1 ^{1}F_{3})$ is
obtained as 55.40MeV. The partial decay widths predicted in this
paper are comparable with the values predicted in Ref.\cite{8}.
\\We have also studied the decay behavior of strange
partners of $D_{2}^{*}(3000)$ and $D_{3}(3099)$ charm states i.e.
$(D_{s2}^{*},D_{s3}) = (2^{+},3^{+})_{\frac{5}{2}}$ with n=1 and
L=3. Masses for these strange charm states are taken as
$3220.66\pm9$ MeV and $3232.50\pm33$ MeV from the theoretical work
\cite{13,16,17,18}. OZI allowed two body strong decay channels of
these two states are also listed in Table \ref{tab:width1f}. For
$D_{s2}^{*}$ state, we observe, $D^{0}K^{-}$ to be the dominant
decay mode with branching fraction $25.94\%$ and for $D_{s3}$ state,
$D^{*0}K^{-}$ to be the dominant mode with branching fraction
$35.95\%$. These strange states also decays to P-wave charm meson
states, but due to small phase space, these modes are suppressed in
our study. Using above $g_{ZH}$, the total decay width for
$D_{s2}^{*}$ comes out to be 178.79 MeV and for $D_{s3}$ it is
120.66 MeV. Taking sum of the partial decay widths to be the total
decay width for these strange states, $D_{s2}^{*}$ state is observed
to be a broader state as compared to its spin partner $D_{s3}$.

\begin{table*}
{\normalsize
\renewcommand{\arraystretch}{1.0}
\tabcolsep 0.2cm \caption{\label{tab:two}Strong decay width of
$D_{2}^{*}(3000)$ with the $J^{P}$ assignment as
$1F_{\frac{5}{2}}(2^{+})$ and $2P_{\frac{3}{2}}(2^{+})$. Ratio
represents $\widehat{{\bf \Gamma}}=
\frac{\Gamma}{\Gamma(D_{2}^{*}(3000) \rightarrow D^{*+}\pi^{-})}$
for $D_{2}^{*}(3000)$. Fraction gives the percentage of the
particular decay width with respect to the total decay width.}
\begin{center}
\begin{tabular}{c|c|c|c|c|c}
\hline \hline $nLs_{l}J^{P}$&Decay channel&Decay
Width(MeV)&Ratio&Fraction&Experimental Value(MeV)\\
\hline
$1F_{5/2}(2^{+})$&$D^{*+}\pi^{-}$&1046.53$g^{2}_{ZH}$&1&13.60&\\
&$D^{*+}\pi^{0}$&531.26$g^{2}_{ZH}$&0.50&6.90&\\
&$D^{*+}\eta$&109.14$g^{2}_{ZH}$&0.10&1.41&\\
&$D^{*+}_{s}K^{-}$&422.87$g^{2}_{ZH}$&0.40&5.49&\\
&$D^{+}\pi^{-}$&2630.35$g^{2}_{ZH}$&2.51&34.20&\\
&$D^{+}\pi^{0}$&1338.14$g^{2}_{ZH}$&1.27&17.39&\\
&$D^{+}\eta$&307.35$g^{2}_{ZH}$&0.29&3.99&\\
&$D^{+}_{s}K^{-}$&1304.87$g^{2}_{ZH}$&1.24&16.96&\\
&Total&&&&$186\pm38$\\
 \hline
 $2P_{3/2}(2^{+})$&$D^{*+}\pi^{-}$&4075.15$\widetilde{g}^{2}_{TH}$&1&24.69&\\
&$D^{*+}\pi^{0}$&2060.89$\widetilde{g}^{2}_{TH}$&0.50&12.48&\\
&$D^{*+}\eta$&387.99$\widetilde{g}^{2}_{TH}$&0.09&2.35&\\
&$D^{*+}_{s}K^{-}$&1754.17$\widetilde{g}^{2}_{TH}$&0.43&10.62&\\
&$D^{+}\pi^{-}$&1952.32$\widetilde{g}^{2}_{TH}$&0.94&23.36&\\
&$D^{+}\pi^{0}$&3856.13$\widetilde{g}^{2}_{TH}$&0.47&11.83&\\
&$D^{+}\eta$&413.76$\widetilde{g}^{2}_{TH}$&0.10&2.50&\\
&$D^{+}_{s}K^{-}$&2002.65$\widetilde{g}^{2}_{TH}$&0.49&12.13&\\
&Total&&&&$186\pm38$\\
\hline \hline
\end{tabular}
\end{center}}
\end{table*}

\begin{figure*}

  \subfloat[Fig 1 Ratio for $2P(2^{+})$
  state]{\includegraphics[width=0.3\textheight,clip=true,angle=0]{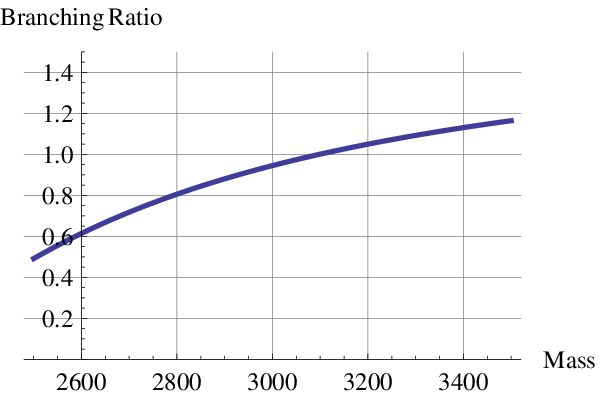}{ \label{ fig:sub1}}}
\hfill
 \subfloat[Fig 2 Ratio for $1F(2^{+})$
    state]{\includegraphics[width=0.3\textheight,clip=true,angle=0]{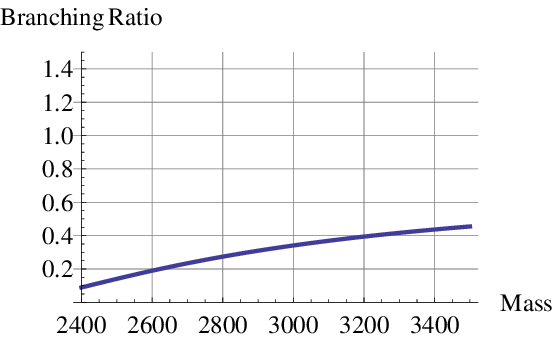}{\label{ fig:sub2} }}

    \caption{Branching ratio
$\Gamma(D_{2}^{*}(3000))\rightarrow\frac{D^{*}\pi}{D\pi}$ for two
possible $J^{P}$'s for $D_{2}^{*}(3000)$ state} \label{fig:test}
\end{figure*}

\begin{table*}
{\normalsize
\renewcommand{\arraystretch}{1.0}
\tabcolsep 0.2cm \caption{\label{tab:width1f}Strong decay width of
$D(1 ^{1}F_{3})$, $D_{s}(1 ^{1}F_{3})$ and $D_{s}(1 ^{3}F_{2})$
charm mesons being the spin and strange partners of $1F(2^{+})$.
Ratio depicts the value $\widehat{{\bf \Gamma}}=
\frac{\Gamma}{\Gamma(D_{J}^{*} \rightarrow D^{*+}\pi^{-})}$ for $D(1
^{1}F_{3})$ and $\widehat{{\bf \Gamma}}=
\frac{\Gamma}{\Gamma(D_{sJ}^{*} \rightarrow D^{*0}K^{+})}$ for
$D_{s}(1 ^{1}F_{3})$ and $D_{s}(1 ^{3}F_{2})$. Last column gives the
branching fraction for these states.}
\begin{center}
\begin{tabular}{c|c|c|c|c}
\hline\hline $nLs_{l}J^{P}$&Decay channel&Decay Width(MeV)&Ratio&Branching Fraction\\
$1F_{5/2}(3^{+})$&$D^{*+}\pi^{-}$&29.03&1&51.84\\
&$D^{*+}\pi^{0}$&14.78&0.50&26.38\\
&$D^{*+}\eta$&2.57&0.09&4.75\\
&$D^{*+}_{s}K^{-}$&9.00&0.32&17.01\\
&Total&55.40&-&100\\
 \hline
$1F_{s5/2}(3^{+})$&$D^{*+}K^{0}$&42.41&0.97&35.15\\
&$D^{*0}K^{+}$&43.38&1&35.95\\
&$D^{*+}\eta$&14.81&0.34&12.27\\
&$D^{*+}_{s}\pi^{0}$&20.04&0.46&16.61\\
&Total&120.66&-&100\\
\hline
$1F_{s5/2}(2^{+})$&$D^{*+}K^{0}$&16.61&0.97&9.29\\
&$D^{*0}K^{+}$&17.00&1&9.50\\
&$D^{*+}\eta$&5.78&0.34&3.23\\
&$D^{*+}_{s}\pi^{0}$&7.86&0.46&4.39\\
&$D^{+}K^{0}$&45.30&2.66&25.33\\
&$D^{0}K^{+}$&46.37&2.72&25.94\\
&$D^{+}\eta$&19.37&1.08&10.29\\
&$D^{+}_{s}\pi^{0}$&21.47&1.26&12.01\\
&Total&178.79&-&100\\
\hline\hline
\end{tabular}
\end{center}}
\end{table*}

\section{Conclusion}
In the present article, we have examined the charm states
$D^{*}_{J}(2460)$, $D_{J}(2560)$,  $D^{*}_{J}(2680)$, $D_{J}(2740)$,
$D^{*}_{J}(2760)$, $D_{J}(3000)$ and $D^{*}_{J}(3000)$ with $J^{P}$
states $1P_{\frac{3}{2}}2^{+}$, $2S_{\frac{1}{2}}0^{-}$,
$2S_{\frac{1}{2}}1^{-}$, $1D_{\frac{5}{2}}2^{-}$,
$1D_{\frac{5}{2}}3^{-}$, $2P_{\frac{1}{2}}1^{+}$ and
$2P_{\frac{1}{2}}0^{+}$ respectively. Here we have used the HQET
lagrangian at the leading order approximation, and studied their two
body strong decay behavior with the emission of light pseudo-scalar
mesons $(\pi,\eta,K)$. We have computed the branching ratios and the
coupling constants $g_{TH}$, $\widetilde{g}_{HH}$, $g_{YH}$,
$\widetilde{g}_{SH}$ for the above states, that can be useful for
the future experimental data to compare with.

Along with this, we have also tentatively identified the $J^{P}$ for
$D_{2}^{*}(3000)$ charm meson which is recently observed by the LHCb
in 2016 \cite{1}. We studied the branching ratio for this state and
concluded its $J^{P}$ to be $1F_{\frac{5}{2}}2^{+}$, and
correspondingly obtained the coupling constant $g_{ZH} \simeq 0.15$.
The obtained coupling constant helps in calculating the strong decay
channels for the experimentally missing $D(1 ^{1}F_{3})$, $D_{S}(1
^{1}F_{3})$ and $D_{S}(1 ^{3}F_{2})$ states. Thus, the observation
of $D_{2}^{*}(3000)$ as $1F_{\frac{5}{2}}2^{+}$ has opened a window
to investigate the higher excitations of charm mesons at the LHCb,
BaBar, BESIII.

\section{Acknowledgement}
The authors gratefully acknowledge the financial support by the
Department of Science and Technology (SB/FTP/PS-037/2014), New
Delhi.

\end{document}